\documentclass{article} 
\usepackage{PRIMEarxiv}
\usepackage{url,hyperref,lineno,microtype,subcaption}
\usepackage[onehalfspacing]{setspace}

\usepackage{algorithm}
\usepackage{algpseudocode}
\usepackage{multirow}
\usepackage{tabularx} 
\usepackage{makecell}
\usepackage{graphicx}
\usepackage{booktabs}
\usepackage{svg}
\usepackage{amsmath}
\usepackage{txfonts}
\usepackage{ragged2e}
\usepackage[numbers]{natbib}
\usepackage[utf8]{inputenc} 
\usepackage[T1]{fontenc}    
\usepackage{hyperref}       
\usepackage{url}            
\usepackage{booktabs}       
\usepackage{amsfonts}       
\usepackage{nicefrac}       
\usepackage{microtype}      
\usepackage{graphicx}

\def\Authors{Shin Sano\,$^{1,3*}$ and Seiji Yamada\,$^{1,2}$}
\def\Address{$^{1}$Department of Informatics, School of Multidisciplinary Sciences,\\ Graduate University for Advanced Studies (SOKENDAI), Hayama , Japan \\
$^{2}$Digital Content and Media Sciences Research Division, National Institute of Informatics , Tokyo, Japan \\ $^{3}$Institute for Creative Integration , Oakland, California, USA}

\begin{document}

\renewcommand\theadalign{bc}
\renewcommand\theadfont{\bfseries}
\renewcommand\theadgape{\Gape[4pt]}
\renewcommand\cellgape{\Gape[4pt]}
\algnewcommand{\IIf}[1]{\State\algorithmicif\ #1\ \algorithmicthen}
\algnewcommand{\endIIf}{\unskip\ \algorithmicend\ \algorithmicif}
\algnewcommand{\FFor}[1]{\State\algorithmicfor\ #1\ \algorithmicdo}
\algnewcommand{\endFFor}{\unskip\ \algorithmicend\ \algorithmicfor}

\pagestyle{fancy}
\thispagestyle{empty}
\rhead{ \textit{ }} 
\fancyhead[LO]{Character Space Construction}  

\title{AI-Assisted Design Concept Exploration \\Through Character Space Construction}

\author{
\Authors{}
\\
\\
\Address{} 
}
\maketitle


\begin{abstract}

 We propose an AI-assisted design concept exploration tool, the ``Character Space Construction'' (``CSC''). Concept designers explore and articulate the target product aesthetics and semantics in language, which is expressed using ``Design Concept Phrases'' (``DCPs''), that is, compound adjective phrases, and contrasting terms that convey what are not their target design concepts.  Designers often utilize this dichotomy technique to communicate the nature of their aesthetic and semantic design concepts with stakeholders, especially in an early design development phase.  The CSC assists this designers' cognitive activity by constructing a ``Character Space'' (``CS''), which is a semantic quadrant system, in a structured manner. A CS created by designers with the assistance of the CSC enables them to discern and explain their design concepts in contrast with opposing terms. These terms in a CS are retrieved and combined in the CSC by using a knowledge graph. The CSC presents terms and phrases as lists of candidates to users from which users will choose in order to define the target design concept, which is then visualized in a CS. The participants in our experiment, who were in the ``arts and design'' profession, were given two conditions under which to create DCPs and explain them. One group created and explained the DCPs with the assistance of the proposed CSC, and the other did the same task without this assistance, given the freedom to use any publicly available web search tools instead. The result showed that the group assisted by the CSC indicated their tasks were supported significantly better, especially in exploration, as measured by the Creativity Support Index (CSI).
\end{abstract}


\section{Introduction}
Our research is motivated by an observation in the professional industrial design domain and a research question derived from it: ``could concept designers' creative activities, especially verbalizing novel design concept phrase, be modeled and computationally supported?'' The concept design used in industrial design is primarily concerned with developing the aesthetics and semantics of products\cite{krippendorff2005semantic}. The visual appearance and meaning behind a product is frequently a key attribute of a product's appeal to consumers \cite{bloch2003individual,alcaide20203d,han2021exploration}, as the functionality of a product is increasingly taken for granted, and users are looking for different levels of appreciation \cite{demirbilek2003product}. Consequently, designers are in charge of not only creating the visual appearance of products but also verbally articulating the product semantics and emotional feelings attached to them \cite{dong2005latent}. Design concepts articulated using this verbal mode often have to be explored, communicated, understood, and approved by the project stakeholders during a design project. \citet{krippendorff2005semantic} called this practice a ``design discourse'', describing, ``Designers have to justify the aesthetics of mass-produced products, products that would ideally be of use to everyone, not individual works of art.''  This characterizes the special role industrial designers play, as distinguished from engineering designers, and this role is most clearly observable in the automobile industry \cite{tovey1992intuitive}.
 
While emotional components in industrial design are important \cite{desmet2002basis,mcdonagh2002visual,norman2004emotional}, automotive customers especially value emotional experiences when they own and use products, such as those related to their self-image and brand messages reflected in vehicles \cite{hekkert2006design,helander2013emotional}. Designers and other stakeholders exchange views on the form of a vehicle design as part of the design process. \citet{tovey1992intuitive} characterized the language they use as ``idiosyncratic and atypical,'' requiring them to explore and communicate different shapes and features in words, such as ``slippery,'' ``exciting,'' ``fluid,'' ``tailored,'' and ``sheer.'' Such language used in car design studios describes particular forms or connotes a ``feeling.'' \citet{bouchard2005nature} described these verbal expressions as ``intermediate representations'' that will eventually be translated into visual aesthetic and semantic features of product designs. Designers in practice often feel that they run out of these words to express different design concepts in different projects.

The present research attempts to uncover professional concept designers' cognitive activities, that are traditionally highly empirical and undocumented, for when they explore and form such verbal representations of design concepts and also attempts to computationally support the process by an AI algorithm. The contributions of this research are as follows:
\begin{itemize}
    \item We defined concept designers’ cognitive activities in exploring, generating, and explaining a design concept with a ``Design Concept Phrase'' (``DCP'') and contrasting concepts, and modeled this process as constructing what we call a “Character Space” (``CS'').
    \item We implemented this model of constructing a CS into an AI algorithm and created an interactive system, which supports concept designers’ creative process described above. 
    \item We defined a DCP as a compound phrase using an adjective-``noun-form-of'' adjective formula.
\end{itemize}

In this paper, we will first discuss the background of the topic space together with related works. We will highlight how professional concept designers utilize a verbal mode of exploration to generate and communicate novel design concepts in the early stage of the design process, especially with an emphasis on their process of using a dichotomy technique, taking as an example the development of an automotive design concept. Then, we propose an AI-assisted interactive system, ``Character Space Construction'' (``CSC''), and its methodologies, utilizing the quadrant system, that is, the CS, for designers to generate and use DCPs. Last, we introduce our experiment with participants in design profession to measure the effectiveness of the tool, followed by discussions.

We illustrate a case study of how they use DCPs from an automotive design, however, these adjective phrases are used in the practice of industrial design in general \cite{shieh2011affective}, fashion design\cite{choo2003effect}, furniture design \cite{khalaj2014comparison}, or even in the tourism industry \cite{duran2019adjectives}.  Therefore, the proposed methods can contribute to various design and marketing fields.

\section{Background}
\begin{figure}[t]
  \centering
  \includegraphics[height=5cm]{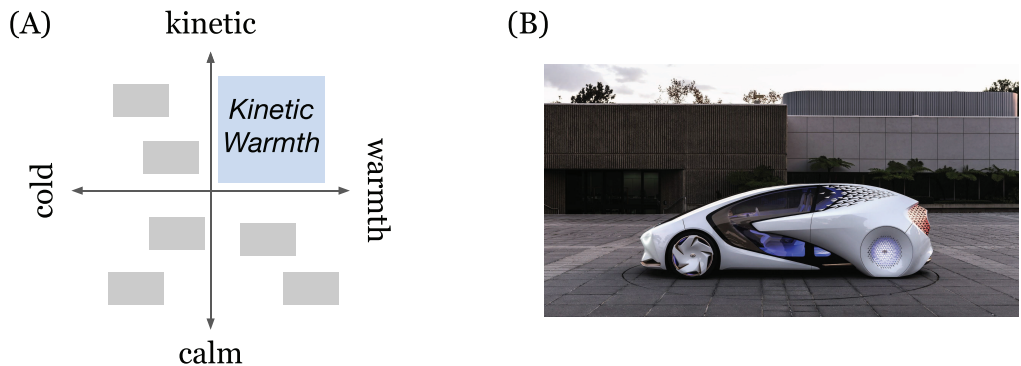}
 \caption{(A) left: Character space (CS) for ``kinetic warmth'' (B) right: Toyota Concept-i design derived from ``kinetic warmth'' (\textcopyright 2021 Toyota Motor Sales, U.S.A., Inc.).}
\end{figure}
\subsection{Defining Design Concept Phrase and Character Space}
First, we define the ``Design Concept Phrase'' (``DCP'') as a compound adjective phrase that conveys product aesthetics and semantics. In the industry, such phrases are sometimes called ``keywords'' \cite{nagai2003experimental}; however, we call them DCP in this research since the term ``keywords'' is used differently in the natural language processing (NLP) context. A DCP, ``kinetic warmth,'' for example, was created by the designers at Toyota's North American design studio for Concept-i (Fig. 1-(B)). Note that the noun ``warmth'' was converted into a noun from the adjective ``warm.'' Ian Cartabiano, the chief designer of Concept-i at Toyota's Calty Design Research studio explained, \textit{``There's a school of design at the moment which makes a car feel just like a laptop on wheels, an impersonal transport unit. We wanted to do something more soulful, intriguing, lively, and human.''} When they identified ``kinetic'' and ``warmth,'' the design team explored other terms, such as ``dynamic,'' ``engaging,'' ``intriguing,'' ``soulful,'' ``human,'' ``lively,'' and ``passionate'' \footnote{Car Design News (2017). CES 2017: Toyota Concept-i in depth. https://www.cardesignnews.com/cdn-live/ces-2017-toyota-concept-i-in-depth/22991.article [Accessed October 22, 2021]. }. Notice that Cartabiano expressed a type of aesthetics and semantics to be avoided, that is, ``a laptop on wheels, an impersonal transport unit.'' They later put those feelings into two other contrasting terms, ``calm'' and ``cold,'' and placed them onto a quadrant system, which we define in this research as the ``Character Space'' (``CS'', Fig. 1(A)). These contrasting terms make ``kinetic warmth'' understanable by stating what is NOT ``kinetic warmth.'' 

A CS visually resembles a dimensionality reduction on 2D semantic space employing multidimensional scaling (MDS); however, the way a CS is formed in practice fundamentally differs from how an MDS is formed. A CS is informally utilized as a synthesis tool employing a quadrant system, on which a user identifies each end of the axis through creative exploration and speculation. The primary focus is the upper-right quadrant, which represents the target concept expressed by a compound adjective; then, the other quadrants will be defined as the user identifies opposite terms to each of the compound adjectives (e.g., ``kinetic-calm'' and ``warm-cold''). These polar terms on the axes consequently form semantic differentials \cite{osgood1957measurement}, however in a CS, the two terms in an orthogonal relationship (e.g., ``kinetic'' on the vertical axis and ``warm'' on the horizontal axis) are identified first before a pair of polar terms on a semantic differential is defined. Also, that these polar terms can be defined by the designers as a part of creation depending on what aspect of a particular adjective they want to emphasize. For instance, an adjective, ``beautiful,'' can be perceived differently depending on the term at the other end (e.g., ``beautiful''$\leftrightarrow$``ugly'' or ``beautiful''$\leftrightarrow$``practical''). A CS is completed when all four ends on the axes are defined. In this research, we label the top end of the vertical axis as word 1 ($w_1$), and the rest are then labeled in clockwise order as word 2 ($w_2$), word 3 ($w_3$), and word 4 ($w_4$). The upper-right quadrant, represented by the combination of $w_1$ and $w_2$, is the target design-concept phrase. All the other quadrants, represented by $w_2$-$w_3$, $w_3$-$w_4$, and $w_4$-$w_1$, are contrasting concepts to be used to explain the target DCP in terms of how and by which attributes they differ and what already exists or what is to be avoided. 

Meanwhile, a multidimensional scaling with MDSs has been widely used as an analytical tool since first being introduced by  \citet{ramsay1975solving} and became a popular data visualization tool when \citet{zimet1988multidimensional} used them in the form of the Multidimensional Scale of Perceived Social Support (MSPSS).  MDSs have been successfully implemented in new product development domains, including concept mapping \cite{trochim1989introduction} and kansei engineering \cite{nagamachi1995kansei}. They visualize the spatial distance between individual cases of a data set according to similarities. The usage of MDSs first involves collecting data on the basis of ratings of statements or semantic differentials for each case, followed by dimensionality reduction done by using statistical techniques, such as principal component analysis (PCA), to yield a scatter plot visualization of the cases on a two-dimensional space. After a visualization with an MDS users usually have to define the labels of the axes by interpreting the collective meanings of the distantly-plotted clusters on an MDS. Therefore, a dimensionality reduction is a methodologically different process from the CS, which we attempt to define in this research.

\subsection{Compound adjective phrase}
While our goal is to address product aesthetics and semantics that evoke end users' emotional feelings attached to them, we focused on adjectives that describe the emotional aspects of product designs.  \citet{hsiao2006fundamental} conducted surveys using product images and adjectives on semantic differentials. The factor analyses identified four main dimensions of the participants' emotional responses to product shapes, characterized by the pairs of polar affective adjectives. They demonstrated the role of adjectives in product designs, yet their focus was not to explore a variety of nuanced adjectives to describe a unique design concept.  

In order to describe distinct quality of a design aesthetics and semantics, we chose to use adjective-adjective combinations. This is called a compound adjective in a coordinated sequence - a sequence in which two or more adjectives severally restrict the head noun (e.g., a tall thin woman). In this example, the phrase therefore means a ``woman who is both tall and thin.'' The criteria for being coordinated and in a sequence is that the adjectives in a coordinated sequence may, but need not, be linked by ``and.'' Also, a change in the order of adjectives in a coordinated sequence may result in a shifting of the emphasis, or even in a stylistic flaw, but will not change the meaning of the phrase \cite{sopher1962sequence}. In addition, the second adjective takes the noun form (e.g., ``kinetic warm''$\rightarrow$``kinetic warmth,'' ``effortless elegant''$\rightarrow$``effortless elegance,'' ``fluid beautiful''$\rightarrow$``fluid beauty'') in order for the phrases to sound complete by themselves. In the case of ``kinetic warm(th),'' the complete phrase would be ``kinetic and warm style'' or ``kinetic and warm car''; however, since the head nouns ``style'' or ``car'' are obvious in the context, they are omitted.  Therefore, the type of phrase we use in this research can be defined as a compound adjective (adjective + adjective) phrase in a coordinated sequence, which is modified to an adjective-noun form. This type of compound phrase is underexplored in the research topic called ``combinational creativity.''

Combinational creativity attempts to produce new ideas by combining existing ideas in unfamiliar ways \cite{boden2001creativity}. A number of studies have identified the effect of combinational creativity as a tool for generating creative design concepts \cite{han2020computational, Georgiev2010d,Chiu2012g}. \citet{nagai2009concept}, in a collaboration with designers, suggested that a new concept can be created by synthesizing two ideas in noun-noun combination phrases. \citet{Chiu2012g} used oppositely and similarly related pairs of words as stimuli for design-concept generation and observed that oppositely related verb-verb combination stimuli could increase concept creativity. The Combinator \cite{han2016combinator} imitates the way the human brain achieves combinational creativity and suggests combined ideas in both visual and textual representations. In these works, combinational creativity focused either on combined product categories or objects derived from noun-noun combinations (e.g., ``desk + elevator'' or ``pen + ruler'') or functional features derived from verb-verb combinations (e.g., ``fill + insert''), and they did not address product aesthetics. 

\subsection{Lexico-semantic exploration tool}
In constructing a CS in our system, we use the ConceptNet knowledge graph \cite{speer2017conceptnet} for exploring compound phrases. ConceptNet is a knowledge graph that connects words and phrases with labeled edges. It is designed to represent varieties of general knowledge to allow users to make word associations and make sense of the meanings behind them. Its knowledge is collected from a wide range of sources, such as expert-created resources, crowd-sourcing, and games with a purpose. Among creativity support tools, \textit{Spinneret} \cite{bae2020spinneret} uses ConceptNet and provides idea suggestions as nodes on a graph. It aims to support divergent thinking, and it explores ideas and provides ``suggestions'' to the user to add to a mind map. \textit{Mini-Map} \cite{chen2019mini} also uses ConceptNet, and it gamifies collaborations with an intelligent agent in creating a mind map.  The both studies chose the ConceptNet as to be suitable for enhancing explorations and mitigating fixation during an ideation process.

\subsection{Metrics of creativity support tools}
Unlike productivity support tools that can be evaluated on the basis of objective metrics, the evaluation of creativity support tools lacks obvious metrics, and there is no one-size-fits-all approach \cite{hewett2005creativity}. In one of the several dimensions of creativity that \citet{sternberg1999handbook} discussed, creativity can be a property of people, a property of products, and a property of a set of cognitive processes. With respect to a property of products, ``originality'' and ``effectiveness'' are popular criteria \cite{barron1955disposition,stein1953creativity,runco2012standard}. Both terms have versions of labels, such as ``novelty,'' ``unusuality,'' or ``uniqueness'' for the former and ``usefulness,'' ``practicality,'' ``appropriateness,'' or ``relevancy'' for the latter \cite{bruner1962conditions,cropley1967creativity,kneller1965art}. They are measured either on the basis of self-report evaluation or rated by experts with the precaution of the inter-rater disagreement.  With respect to evaluating a property of people, it tends to lead to a concern with individual differences between people. 

In this research, we used the Creativity Support Index (CSI)\cite{cherry2014quantifying}, which covers the two properties of creativity above, that are, a property of products, and a property of a set of cognitive processes, both in terms of users' own perceptions.  This would be suitable for the present research, especially when the objective is to cater a tool, which is designed with certain functional requirements in mind, to support the effectiveness of designers' cognitive processes.  We will discuss the further details of the CSI, its benefits and application in section 4.3.

\section{Method}
\subsection{Preliminary study for adjective-adjective compound phrases}
As discussed in section 2.2, we intend to use compound adjective phrases for the DCP.   Despite the fact that some related works demonstrated the role of adjectives in conveying users' emotional responses to product aesthetics\cite{hsiao2006fundamental}, or the effectiveness of combinational creativity\cite{hsiao2006fundamental}, this particular type of combinational phrase is under-researched in the context of supporting designers' creativity, there is no prior work that demonstrated the effectiveness of adjective-adjective compound phrase compare to different types with other parts-of-speech(PoS), such as noun-noun or verb-verb. Thus, we conducted a preliminary experiment comparing phrases with different PoS to confirm that adjective phrase combinations are suitable for communicating product aesthetics. 

We used Survey Monkey to recruit 55 participants whose job function is ``arts and design.'' Six sample phrases were randomly generated for each of four different combinations of PoS: adjective-adjective (AA), adjective-noun (AN), noun-noun (NN), and verb-verb (VV). For the selection of these words, we randomly extracted the words for all combinations from a corpus, which we created with English automotive design articles containing about 1.4 million words on Sketch Engine \cite{kilgarriff2004itri}. The following filtration was performed equally for all PoS combinations. First, we removed non-English terms and proper nouns and then removed words whose relative frequency per a million tokens was less than 1.0, compared with enTenTen15 \cite{jakubivcek2013tenten}, a large web text corpus containing about 1.6 billion words. This made the list of all words, regardless of PoS, more suitable for designing concept expressions that participants can readily recognize and make a judgment about. The participants rated 24 randomly generated two-word combinational phrases in randomized order on a 7-point Likert scale on the basis of the degree to which they agreed with the statement ``I can imagine the product aesthetics, characters, mood, and emotional quality that are conveyed by the product design.'' An ANOVA-Tukey test identified homogeneous subsets as [AA/AN/NN] and [VV/NN], where the mean score of the former was significantly higher $(p =0.05^*)$. Therefore, using adjective-``noun-form-of'' adjective phrases, as opposed to using noun-noun or verb-verb phrases, would yield optimized effect on addressing product aesthetics and semantics, as well as emotional qualities attached to them we are trying to accomplish in this research.

Further details on the individual comparisons are as follows. There was a significant difference $(p =.038^*)$ in mean scores between AA $(3.890, \sigma =1.121)$ and VV $(3.300, \sigma =1.205)$, a significant difference $(p =.040^*)$ between AN $(3.885, \sigma =1.025)$ and VV, no significant difference $(p =.581)$ between AA and NN $(3.612, \sigma =1.232)$, no significant difference $(p =.599)$ between AN and NN, no significant difference $(p =.485)$ between NN and VV, and no significant difference $(p =1.000)$ between AA and AN. 

\subsection{System overview}

\begin{figure}[t]
  \centering
  \includegraphics[width=\linewidth]{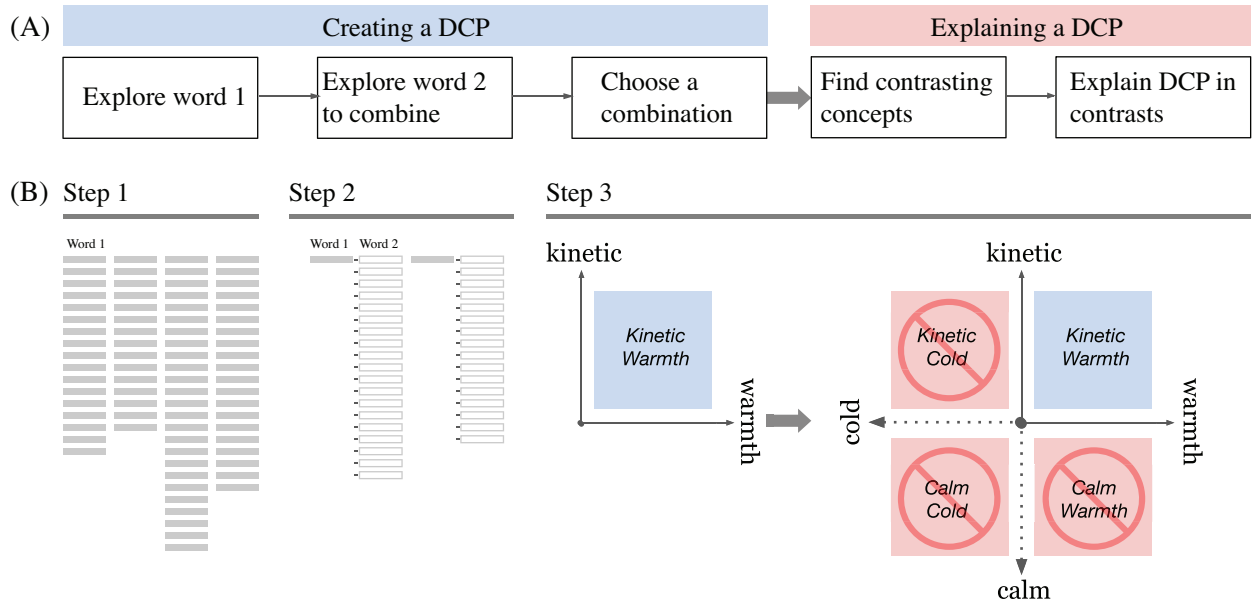}
  \caption{A Process forming and explaining a DCP (A), which is modeled as constructing a CS (B) }
\end{figure}
\begin{figure}[t]
  \centering
  \includegraphics[height=5cm]{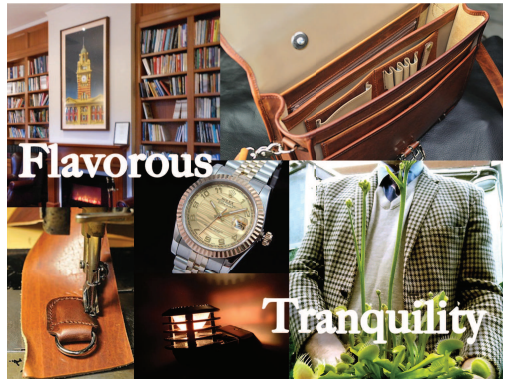}
  \caption{An example of a moodboard, which is inspired by a DCP ``flavorous tranquility''}
\end{figure}
The present system, Character Space Construction (CSC), is an AI-assisted interactive system. A new contribution of the CSC is that it helps designers to construct a 2-dimensional CS as a quadrant system, which yields a DCP, a compound adjective phrase, that portrays a design concept and explains it in semantic comparisons. The CSC is a web application that has all the functions above in one system and is operated in an interactive manner. The algorithm of the system, which constructs a CS by letting the user set four words ($w_1$-$w_4$), will be discussed in detail in section 3.3. The CSC application consists of a front-end web application written in JavaScript, HTML, and CSS and a back-end web server written in Python with a MySQL database hosted on PythonAnywhere, which is a hosting service for Python web applications. The MySQL database was built on PythonAnywhere and holds the data of word embeddings from the ConceptNet Numberbatch \cite{speer2017conceptnet}.

\subsection{Character Space Construction algorithm} 

As discussed in Section 2.1., some concept designers empirically use a CS as a synthesis tool, not as an analytical tool, in a process of forming a DCP. On a quadrant system, two terms in the upper right quadrant are identified first then the opposite term on each vertical and horizontal axis are identified. This unique process has not been formalized in the past to the best of our knowledge. Fig. 2(A) illustrates the sequence of designers' cognitive activities.  It starts with searching and selecting a few candidates of the first words of a compound adjective, followed by searching and identifying the second adjective to be combined with the candidates of word 1.  Once word 1 and word 2 are identified as a DCP, designers then attempt to find contrasting words for each to explain the DCP in comparison to those terms, with or without CS. Fig. 2(B) shows the process model of constructing a CS in sequence, replicating the process in Fig. 2(A) in a structured manner. A formalization of constructing a CS not only clarifies the designers' process of verbal design concept explorations, but also signifies the next phase of creative activities, that is, creating a visual mood board.  Although creating a mood board is outside of this research's scope, it should be noted that this quadrant system on a CS will eventually help them explore the visual images for a target design concept, followed by finding visual images that should be avoided. Fig. 3 is an example of what a mood board would look like with a DCP, ``flavorous tranquility.''\footnote{All the images in Fig. 4 are licensed under Creative Commons in either of the following conditions: CC PDM 1.0, CC BY-NC-ND 2.0, CC BY 2.0, or CC BY-SA 2.0}

Adhering to this model, the CSC tool consists of three steps in a UI (Fig. 4). The top section is step 1, where users can input a design brief in sentence form to start searching for word 1 $w_1$. A design brief is a written description of a project that requires some form of design, containing a project overview, its objectives, tasks, target audience, and expected outcomes \cite{phillips2004creating,koronis2018impact}. In design competitions, design briefs are usually explicitly written material that calls for designers' entries. In industry practice, design briefs may be implicitly formed through discussions between stakeholders in a project. In the top right section above step 2, the system provides a search window in which the participant can input queries in case they did not find any word that they wanted to use in the Explorer, after candidate words are displayed. The lower left space is allocated for ``Explorer,'' in which ranked search results are shown either as $w_1$ or combinations of $w_1$ and $w_2$, depending on the phase of the exploration. 

The middle right section is step 2, where users can choose and pool candidates for $w_1$ and search for candidates for $w_2$ that are combined with the $w_1$ candidates. The users at this point will see a variety of adjective phrases ($w_1$-$w_2$) in the Explorer as ranked lists in columns, grouped by $w_1$.
\begin{figure}[t]
  \centering
  \includegraphics[width=\linewidth]{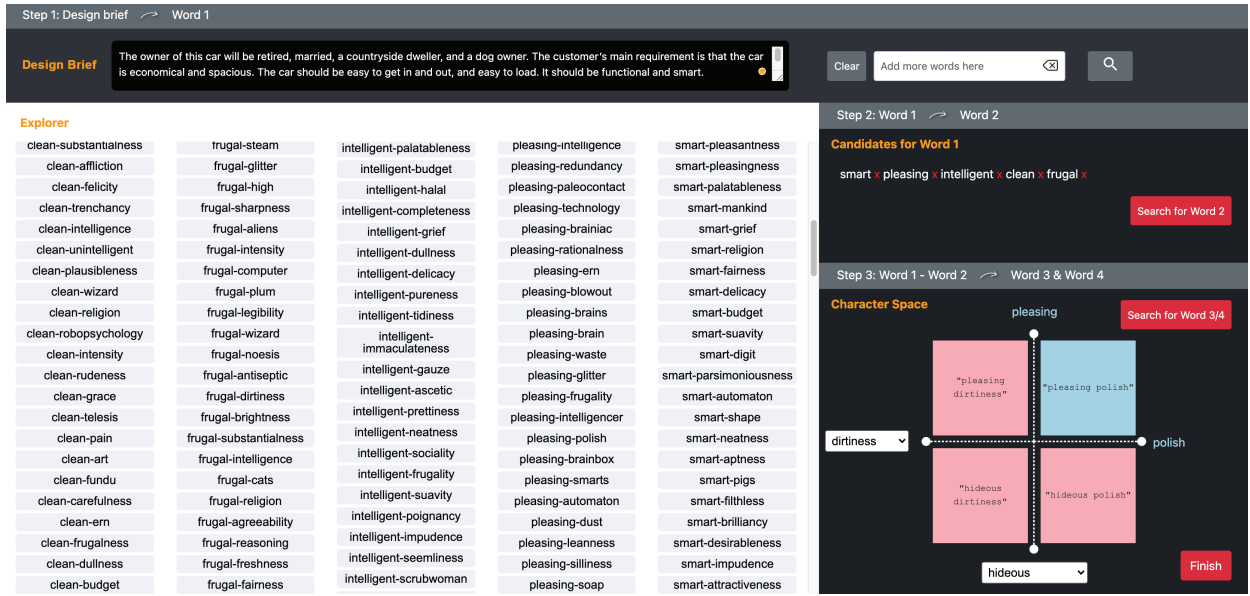}
\caption{Web application user interface of the Character Space Construction (CSC). The top sections are the input windows for ``Design Brief'' and individual query words. The largest section is allocated to ``Explorer'', where candidates for word 1 and word1 - word 2 compound phrases are presented. On the right sidebar, the ``Candidates for Word 1'' section shows the words a user selected and the ``Search for Word 2'' button. ``Character Space'' shows a quadrant system as a user chooses word 1 and word 2, that trigger to show the candidates of Word 3 and Word 4 as opposing concepts.}
 \end{figure}

Finally, the lower right section is the CS, which is a quadrant system with all four words set at the ends of each axis. The first quadrant of the CS is set by dragging and dropping an adjective phrase chosen by the user from the lists in the Explorer. Once the user finalizes the first quadrant of the CS with adjective phrases ($w_1$-$w_2$) and clicks the ``Search for word 3/4'' button, the system will search for antonyms to $w_1$ and $w_2$, as $w_3$ and $w_4$, respectively. The candidates for $w_3$ and $w_4$ are shown in a pull-down menu at the end of each axis so that the user can choose a suitable contrasting word for each $w_1$ and $w_2$ to complete the CS. The upper right quadrant, represented by the combination of $w_1$ and $w_2$, is the target design-concept phrase. All the other quadrants, represented by $w_2$-$w_3$, $w_3$-$w_4$, and $w_4$-$w_1$, are contrasting concepts to be used by the participants to explain the target design concepts in comparison with opposing concepts. After clicking on the ``Finish'' button, the system generates an explanation of the DCP as a template matched sentence, like ``My design concept is $w_1$-$w_2$. It has a sense of $w_2$ yet is $w_1$, not $w_3$. It is $w_1$ but not $w_4$. In this design, $w_1$ and $w_2$ can go together.'' Algorithm 1 illustrates the procedure described above. Table 1 shows brief explanations of each sub-function in Algorithm 1.


\begin{algorithm}
\caption{Character Space Construction}
\begin{algorithmic}[H]
\Function{construct\_character\_space}{}
\State  $\mathit{query\_words}$:= GenerateQuery(design\_brief) 
\State  $w_1\_initial\_\mathit{candidates}$:=[]
\State $\mathit{phrases}$:=[]
    \State $phrase\_scores$:=[]
    \For {each $w$ in $\mathit{query\_words}$}
    \State {$w_1\_\mathit{initial\_candidates}$.append(SearchCandidates($w$))}
    \State sort $w_1\_\mathit{initial\_candidates}$ by CalculateWordScore($w_1$)
    \State display $w_1\_\mathit{initial\_candidates}$ 
    \State  $w_1\_\mathit{candidates}$:= let user {\texttt{set}} 5 words from $w_1\_\mathit{initial\_candidates}$ 
    
    \EndFor
    \For {each $w_1$ in $w_1\_\mathit{candidate}$} 
        \State $w_2\_\mathit{candidates}$:= SearchCandidates($w_1$)
        \For{each $w_2$ in $w_2\_\mathit{candidates}(w_1)$}
            \If {CalculateWordScore($w_2)>=1.7$} 
                \State{$\mathit{phrases}$.append($w_1,w_2$)}
                \State{$\mathit{phrase\_scores}$.append(CalculatePhraseScore($w_1,w_2$))}
            \EndIf 
        \EndFor
    \EndFor
    \State sort $\mathit{phrases}$($w_1$, $w_2$) by $phrase\_scores$
    \State display $\mathit{phrases}$
    \State  ($w_1,w_2$):= let user {\texttt{set}} a phrase from $\mathit{phrases}$ 
    \State display CharacterSpace($w_1,w_2$)
    \State $w_3\_candidates$:= SearchAntomnyms($w_1$) 
    \State $w_4\_candidates$:= SearchAntomnyms($w_2$) 
    \State display CharacterSpace($w_1,w_2, w_3, w_4$)
    \State display GenerateExplanation($w_1,w_2, w_3, w_4$)
\EndFunction
\end{algorithmic}
\end{algorithm}   

\begin{table}[h]
\begin{center}
\caption{List of functions in CSC algorithm}
\vspace*{0.5\baselineskip}
\begin{tabular}{ll} \toprule
Function&Description\\
\midrule
\textbf{GenerateQuery} & Extract adjectives from design brief \\
\textbf{SearchCandidates} & Search related words from ConceptNet\\
\textbf{CalculateWordScore} & Calculate score of word according to usefulness \\
\textbf{CalculatePhraseScore} & Calculate score of adjective phrase according to creativity\\
\textbf{SearchAntonyms} & Search antonyms from ConceptNet\\
\textbf{CharacterSpace} & Draw a quadrant with set words\\
\textbf{GenerateExplanation} & Generate an explanation with template matching\\
\bottomrule
\end{tabular}
\end{center}
\end{table}

For the function \textbf{CalculateWordScore}, the preliminary study gave us an opportunity to observe a list of randomly chosen adjectives. Despite being extracted from the domain-specific corpus we created, the list included a considerable number of unusable adjectives. These unusable adjectives included ones with negative connotations (e.g., amateurish, costly, detrimental), temporality (e.g., first, latest, recent, subsequent), distance (e.g., far, near, close), or utilities (e.g., guest, sourced, takeaway) as well as those that were too concrete in terms of expressing visuals (e.g., square, yellow, golden), to name a few. 

To rule out such unusable adjectives, we used a binary labeling (usable/unusable) for a design concept by five professional automotive designers, who were independent from this research, and we used the count of usable labels and the word embedding vectors from the ConceptNet Numberbatch \cite{speer2017conceptnet} to form a model that predicts the usefulness score of unknown adjectives using gradient boosting \cite{friedman2001greedy}, which is a decision-tree-based ensemble machine learning algorithm.  We used a Python software library called XGBoost (v0.90)\cite{chen2016xgboost}, which uses a gradient boosting framework. We implemented a linear regression predictive model that was trained to minimize the prediction errors (RMSE) in a 10-fold cross validation that yielded the max depth of the decision tree and the learning rate as $6$ and $0.05$, respectively. The average RMSE of the prediction with the best hyperparameters on test data was $1.406$. We used this model to predict usefulness scores for new adjectives, which were used to display the candidate adjectives for $w_1$ in descending order. 

For the function \textbf{CalculatePhraseScore}, we tried to replicate the predictor of the degree of creativity, which is the threshold of the similarity of two words, from the related works on combinational creativity \cite{han2019three, han2018conceptual}, which demonstrated that noun-noun combinations that represent good (award-winning) design concepts fell into a certain range of distance between two nouns. However, our test with five professional automotive designers on adjective phrases in terms of creativity with pre-calculated similarity did not show a clear normal distribution in the scores to determine a threshold. Yet, some brackets for similarity showed better or worse scores than the others, so we assigned these scores to unknown adjective phrases in accordance with the calculated similarity values for each instance. These phrase scores were used to display the candidate phrases for $w_1$-$w_2$ in descending order.

\section{Experiment Design}

\subsection{Participants and independent variables}
The study protocols below have been approved by the Institutional Review Board of the National Institute of Informatics, Tokyo, Japan on February 15th, 2021 (Approval number 0042).

172 participants, whose job function was arts and design and who were fluent in English, were recruited via Prolific, an online user test participant recruiting service. 17 duplicate participants were removed. Of the 155 non-duplicate participants, 15 ($9.68\%$) withdrew due to system trouble, and 37 ($33.55\%$) did not complete the study for unknown reasons. The withdrawal rate was high due to the complex and time-consuming nature of the task. That left us with a total of 103 participants (54 M, 49 F) who completed the study, with a mean age of 28.1 years ($\sigma=8.51$). 

We recruited professional designers who would utilize the tool, instead of recruiting potential end users who would consume the product, because as discussed in Section 2.4, the present research focuses on an evaluation of a creative support tool (CST), most commonly defined as tools to make more people more creative more often \cite{shneiderman2002creativity}.

All of the participants who completed the study used either laptops or desktop computers of their own. The participants who completed the study were paid US\$15. The independent variables differed by whether or not the participants used the CSC system (Fig. 4).  When not using the CSC, the participants were given complete freedom to use any publicly available web search tools, such as Google search, an online thesaurus, or Wikipedia. All of the participants were asked to perform the same task twice with (experiment) and without (control) the CSC system.  The order of the tool they used in two tasks was assigned randomly in a counterbalanced order.  They were given different design briefs (Table 2) for each task to mitigate a learning effect, however, we deemed those two design briefs equivalent in terms of influences on the results for the following rationals.  First, both briefs share the the domain of the products (automotive products), and the basic component of the design briefs: target audience, the context of the product in use, and expected features. Second, the textual stimuli in both design briefs are described in the same abstraction levels.  \citet{gonalves2012find} reported although different visual stimuli strongly provoked different outcomes, difference in textual stimuli did not show significant difference in the outcomes, unless they are substantially different in specificity of domain, level of abstraction, and the distance between the baseline and the supplemental briefs.

We did not offer a blank CS for the control participants because using the CS is a part of our proposed method. The participants in the control group were given instructions and an example of how to generate an explanation of their DCPs. They participated in the experiment online using Survey Monkey, an online platform, to which the participants were redirected from the recruiting service, Prolific. Each participant was given video instructions explaining the goal of the task and how to use the CSC tool (experiment group only) before the task. The CSC tool was given to the experiment group with a URL link.





\begin{table}[h]
\begin{center}
\caption{Two variations of design brief}
\vspace*{0.5\baselineskip}
\begin{tabular}{p{3cm}p{13cm}} 
\toprule
Design brief A:&
The owner of this car will be retired, married, a countryside dweller, and a dog owner. The customer's main requirement is that the car be economical and spacious. The car should be easy to get in and out of and easy to load. It should be functional and smart.
\\ \midrule
Design brief B:&
Imagine a car for a family with small kids in a suburban community of 2030, where technologies are ubiquitous. Consider three important aspects of the car: harmony with local charm, fun activities, and safety.\\
\bottomrule
\end{tabular} 
\end{center}
\end{table}

\subsection{Stimuli and tasks}
After reading a brief, the participants were prompted to start the task. Those with the experiment condition (CSC tool) was instructed to start the task by copying and pasting the design brief to the CSC tool, and as described in section 3.3, the CSC system provides a search window in which the participant can input queries in case they did not find any word that they wanted to use in the Explorer. 

For both experiment and control conditions, there were two tasks.
The first task was ``to create a unique and original DCP as an adjective-noun phrase form that will inspire the audience (described in the design brief) to imagine the character of the design.'' ``Effortless elegance'' was given to the participants as an example of an adjective phrase in the instruction video. The second task was to explain their own DCP with contrasting words. As an example, the adjective ``uneasy'' and the noun ``clumsiness'' were presented as contrasting words for ``effortless'' and ``elegance,'' respectively. An example of an explanation sentence was also given as follows: ``My design concept is `effortless elegance.' It has a sense of elegance, yet it is effortless, not uneasy. It is effortless but not clumsy. In this design, effortless and elegance can go together.'' 

For the experiment condition, this explanation was generated automatically using all $w_1$ through $w_4$ using template matching as long as they completed the CSC properly, and the CSC provided a ``copy'' button that allowed the participants to copy the automatically generated explanation to the clipboard so that they could paste it into the survey. However, they were also given the freedom of editing it in the survey questionnaire. The participants with the control condition had to generate and write an explanation down in the survey questionnaire on their own, however, they were also provided exactly the same example paragraph described above.  In the experiment condition. participants were also given a unique session ID by the CSC system so that we could match the system log to examine how each participant interacted with the words and phrases.

\subsection{Evaluation method}
As discussed in section 2.4, we used the CSI \cite{cherry2014quantifying} as a post-task psychometric measurement to compare two conditions, with (experiment) or without (control) the CSC, in terms of creativity support. The CSI enables us to quantify the cognitive processes of users using psychometric scales for six factors: Exploration, Expressiveness, Immersion, Enjoyment, Results Worth Effort, and Collaboration (Table 4). The CSI evaluates a creativity support tool itself, and focuses on the experience of using it to create rather than trying to directly evaluate a property of products.  Instead, the CSI evaluates the result of a creation in relation to an effort a user made, such as ``I was satisfied with what I got out of the system or tool.''  This is suitable for tools designed for experienced users, who know what the creative outcomes are and what the ideal experiences in creation are, as opposed to the tools designed for novice users where creativity is not sufficient and thus needs support. The CSI also offers flexibility, such that it can be applied to various tools and scenarios over time and provides standardized measurement. It is robustly developed as each of the six factors has two different statements (Table 3) that improve the statistical power of a survey, which is deployed in the CSI. In combination with paired factor comparisons for each of the six factors (Table 4), it also provides insight into what aspects of creativity support may need attention. 

The CSI has a rigorous protocol that the researcher should follow in order for the measurement and analysis to be universal and reliable. For instance, our CSC tool is not designed for collaborative tasks; however, the CSI protocol discourages researchers from skipping the statements in the Collaboration factor, still allowing participants to rate the collaboration statements. Instead, the CSI protocol allows adding ``N/A'' responses to statements that belong to the collaboration factor, which we incorporated in our survey.

\begin{table}[h]
\begin{center}
\caption{The 12 Agreement Statements on the CSI. Each agreement statement is answered on a scale of “Highly Disagree” (1) to “Highly Agree” (10). In deployment,the factor names are not shown, and the participant does not see the statements grouped by factor\cite{cherry2014quantifying}.}
\vspace*{0.5\baselineskip}
\begin{tabular}{p{16.7cm}} 

\toprule
\textbf{Collaboration:}\\
\small1. The system or tool allowed other people to work with me easily.\\
\small2. It was really easy to share ideas and designs with other people inside this system or tool.\\

\midrule
\textbf{Enjoyment:}\\
\small1. I would be happy to use this system or tool on a regular basis.\\
\small2. I enjoyed using the system or tool.\\

\midrule
\textbf{Exploration:}\\
\small1. It was easy for me to explore many different ideas, options, designs,or outcomes, using this system or tool.\\
\small2. The system or tool was helpful in allowing me to track different ideas, outcomes, or possibilities.\\

\midrule
\textbf{Expressiveness:}\\
\small1. I was able to be very creative while doing the activity inside this system or tool.\\
\small2.The system or tool allowed me to be very expressive.\\

\midrule
\textbf{Immersion:}\\
\small1. My attention was fully tuned to the activity, and I forgot about the system or tool that I was using.\\
\small2. I became so absorbed in the activity that I forgot about the system or tool that I was using.\\

\midrule
\textbf{Results Worth Effort:}\\
\small1. I was satisfied with what I got out of the system or tool.\\
\small2. What I was able to produce was worth the effort I had to exert to produce it.\\
\bottomrule
\end{tabular} 
\end{center}
\end{table}
\begin{table}[h]
\begin{center}
\caption{The Paired-Factor Comparison Test has 15 comparisons for each pair, a user will choose a factor description in response to the following statement: “When doing this task, it’s most important that I’m able to...”\cite{cherry2014quantifying}}
\vspace*{0.5\baselineskip}
\begin{tabular}{p{10.7cm}} 
\toprule
1. Be creative and expressive.\\
\midrule
2. Become immersed in the activity.\\
\midrule
3. Enjoy using the system or tool\\
\midrule
4. Explore many different ideas, outcomes, or possibilities.\\
\midrule
5. Produce results that are worth the effort I put in.\\
\midrule
6. Work with other people.\\
\bottomrule
\end{tabular} 
\end{center}
\end{table}

\begin{figure}[t]
  \centering
  \includegraphics[height=3cm]{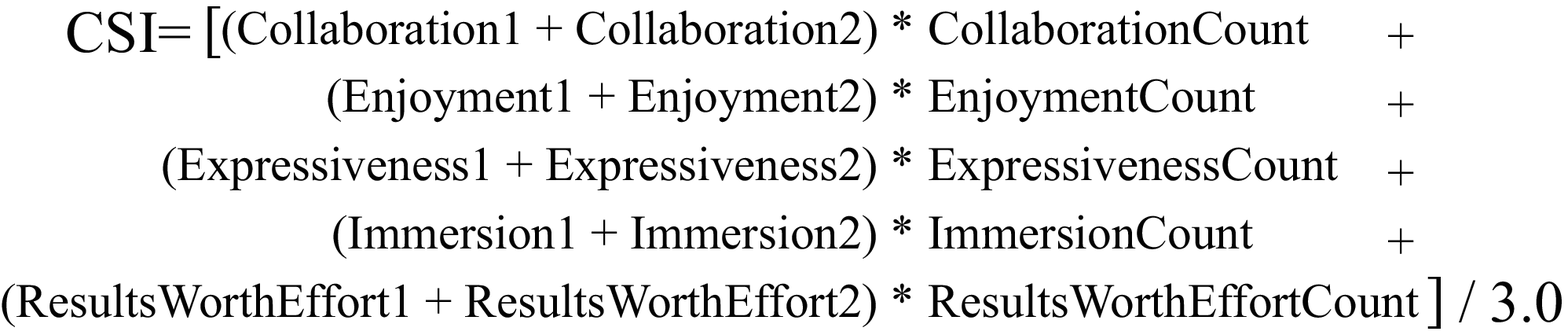}
  \caption{Equation for scoring the CSI \cite{cherry2014quantifying}}
\end{figure}

Also, the final CSI score will be weighted by the scores of paired-factor comparisons, so if a participant does not think of the collaboration factor as important, it is reflected in the final score accordingly. The only other modification we made was to replace ``the system or tool'' with ``The CSC'' in each statement (e.g., ``I enjoyed using the CSC.'') that we were able to do so in. We asked 12 questions on how strongly they agreed or disagreed with each statement on a 10-point Likert scale after each task was completed and 15 paired-factor comparison questions at the end, emphasizing that they were comparing the factors that they thought were important for the tasks, not for the different tools. To check the reliability of each factor, we examined the similarity of the scores (Table 5) across the two different statements.

\subsection{Participant profile}
Since we designed the CSC with professional designers in mind, we were interested in examining the CSI ratings depending on the years of experience and how often the participants performed the particular task we attempted to support with the CSC, that is, generating DCPs especially in the aesthetic sense and explaining them to stakeholders. For the former, we bracketed the years-of-experience groups as: less than 1 year, 1 to 3 years, 3 to 5 years, 5 to 10 years, and 10+ years. For the latter, the groups were divided as: less than quarterly or never, quarterly, monthly, weekly, and daily.


\section{Results}
\subsection{Creativity Support Index} 


\begin{table}[h]
\begin{center}

\caption{CSI and factor scores}
\vspace*{0.5\baselineskip}
\begin{tabular}{lccrrrrc} \toprule

&\makecell{Cronbach's\\Alpha\\per Factor}&\makecell{Average\\Factor\\Count}&\multicolumn{2}{l}{Factor Score} &\multicolumn{3}{l}{Weighted Factor Score}\\							
\midrule
&&&\small{\makecell{CSC\\(N=93)}}&\small{\makecell{Control\\(N=95)}}&\small{\makecell{CSC\\(N=93)}}&\small{\makecell{Control\\(N=95)}}&\\							
&&\makecell{Mean\\($\mathit{\sigma}$)}&\makecell{Mean\\($\mathit{\sigma}$)}&\makecell{Mean\\($\mathit{\sigma}$)}&\makecell{Mean\\($\mathit{\sigma}$)}&\makecell{Mean\\($\mathit{\sigma}$)}&\makecell{$p$\\(\small{Weighted} \\\small{Factor}\\ \small{Score})}\\
\midrule
Collaboration&$0.829$&\makecell{$0.38$\\($0.8$)}&\makecell{$4.96$\\($6.07$)}&\makecell{$5.37$\\($6.53$)}&\makecell{$2.69$\\($8.24$)}&\makecell{$3.06$\\($8.39$)}&$0.758$\\		\midrule										
Enjoyment&$0.894$&\makecell{$2.26$\\($1.36$)}&\makecell{$15.48$\\($4.13$)}&\makecell{$14.11$\\($3.88$)}&\makecell{$35.63$\\($24.84$)}&\makecell{$30.8$\\($20.76$)}&$0.149$\\	\midrule											
Exploration&$0.787$&\makecell{$3.89$\\($0.88$)}&\makecell{$15.87$\\($3.48$)}&\makecell{$13.35$\\($3.94$)}&\makecell{$61.91$\\($20.6$)}&\makecell{$51.32$\\($18.73$)}&$< 0.010^{**}$\\	\midrule											
Expressiveness&$0.843$&\makecell{$3.69$\\($1.04$)}&\makecell{$14.45$\\($3.85$)}&\makecell{$13.63$\\($3.25$)}&\makecell{$52.56$\\($19.71$)}&\makecell{$50.61$\\($19.48$)}&$0.496$\\											\midrule

Immersion&$0.846$&\makecell{$1.78$\\($1.07$)}&\makecell{$11.06$\\($4.76$)}&\makecell{$11.37$\\($4.41$)}&\makecell{$19.95$\\($16.12$)}&\makecell{$21.15$\\($15.7$)}&$0.605$\\	\midrule											
\makecell[l]
{Result Worth\\Effort}&$0.811$&\makecell{$3.01$\\($1.25$)}&\makecell{$14.66$\\($3.75$)}&\makecell{$14.03$\\($3.11$)}&\makecell{$43.35$\\($20.67$)}&\makecell{$42.03$\\($21.9$)}&$0.671$\\													\midrule \midrule												
CSI&&&&&\makecell{72.03\\(16.67)}&\makecell{66.32\\(14.22)}&$0.012^*$\\

\bottomrule
        \end{tabular}
    \end{center}
\end{table}

Table 5 shows the total CSI scores and factor scores between the experiment (CSC) group and the control group. The mean CSI score of the experiment group, which used our proposed CSC ($72.03 ,\sigma = 16.67)$, was significantly higher ($p = .012 < .05^{*}$, Cohen's d $= .369)$ than that of the control group ($66.32, \sigma = 14.22$)(Fig. 6, error bars show standard error). Note that 4 participants out of 103 were excluded due to their DCP being invalid because the compound phrase ($w_1$-$w_2$) was impossible to interpret, hence leaving us with 198 cases from 99 participants. Ten cases were excluded as outliers whose CSI scores were more than two standard deviations away from both sides of the mean, which left us with 188 cases (93 experiments, 95 control) for the final analysis. 

For the reliability of the ratings within each factor, the Cronbach's alpha for Exploration, Expressiveness, Immersion, Enjoyment, Results Worth Effort, and Collaboration was $.787, .843, .846, .894, .811$, and  $.829$, respectively. As for the factors of the CSC the participants found important, we compared the mean factor counts from paired factor comparisons, which were $3.8876 (\sigma = 0.88)$, $3.6854 (\sigma = 1.04)$, $1.7753 (\sigma = 1.07)$, $2.2584$ ($\sigma = 1.36)$, $3.0112 (\sigma = 1.25)$, and $.3820 (\sigma = 0.80)$, respectively. 

Weighted factor scores were calculated by multiplying a participant's factor agreement rating on the statement that belongs to the factor by the factor count (Fig.5) \cite{cherry2014quantifying}. This was done in order to make the weighted factor score more sensitive to the factors that were the most important to the given task. The weighted factor score of Exploration for the experiment group ($61.91$, $\sigma = 20.6$) was significantly higher ($p < .001^{**}$, Cohen's d $= .679)$ than that of the control group ($51.32, \sigma = 18.73$). The other weighted factor scores did not indicate a significant difference between the experiment and control groups. As for the perceived importance of the factors by the participants, the mean count of the factors for Exploration [$3.89$ ($\sigma = 1.04$)] was among the highest, and that of Collaboration [$0.38$ ($\sigma = 0.8$)] was among the lowest.

A partly-repeated two-way ANOVA was performed in order to check the order effects.  The within-participants factor is the tool they used (CSC or control), and the between-participants factor is the order they used the tool (CSC first or control first). We report a significant order effect ($p = .007^{**}, F = 7.70$) between CSC-first or control-first groups.  As for the within-participants test, there were no significant main effect or significant interaction between the tool and the order.

\subsection{CSI score distribution by participants profile}
Fig. 7 shows the distribution of CSI score between the CSI(experiment) group and control group by years of experience brackets in ``arts and design'' profession.  While the years of experience was asked as a categorical choice we examined a Pearson rank correlation between the CSI score and the median rank in each bracket with respect to the years of professional experiences for each group.  Weak correlations were observed between the CSI score and the years of experience for both CSC ($r = .235, p = 0.023^{*}$ and the Control($r = .237, p = 0.021^{*}$ groups. This indicates that the longer the experience, the more they value the benefit of the tool they used, which is consistent in both CSC and control groups.  The mean CSI scores of CSC group for: less than 1 year, 1 to 3 years, 3 to 5 years, 5 to 10 years, and 10+ years were $65.27$, $69.9$0, $69.61$, $76.41$, and $76.87$, respectively. The mean CSI scores of the control group for the same years-of-experience brackets were $61.54$, $63.01$, $65.54$, $69.62$, $71.98$, and $66.32$, respectively. As for how often the participants performed the particular task, we did not observe any such tendencies in the CSI score by the frequency of the DCP creation task they engaged in. Table 6 shows the illustrative examples of randomly selected DCPs generated by the participants per tools and design briefs.


 \begin{figure}[h]
  \centering
  \includegraphics[width=8cm]{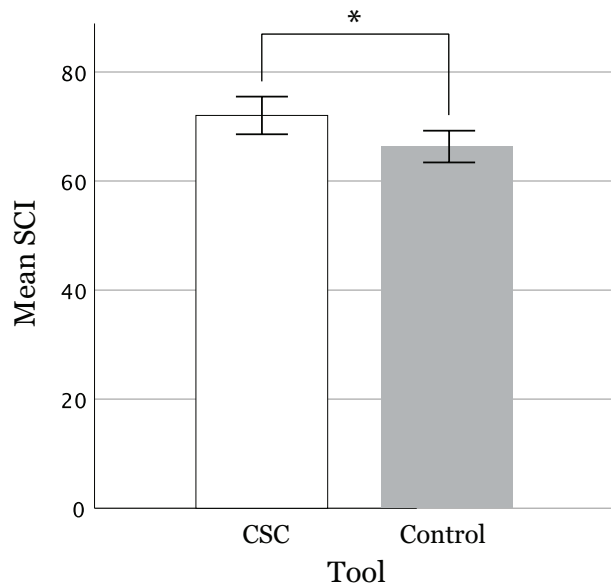}
  \caption{CSI scores between experiment (CSC) group and Control group}
 \end{figure}
 
 \vspace*{1\baselineskip}
 
 \begin{figure}[h]
  \centering
  \includegraphics[width=10cm]{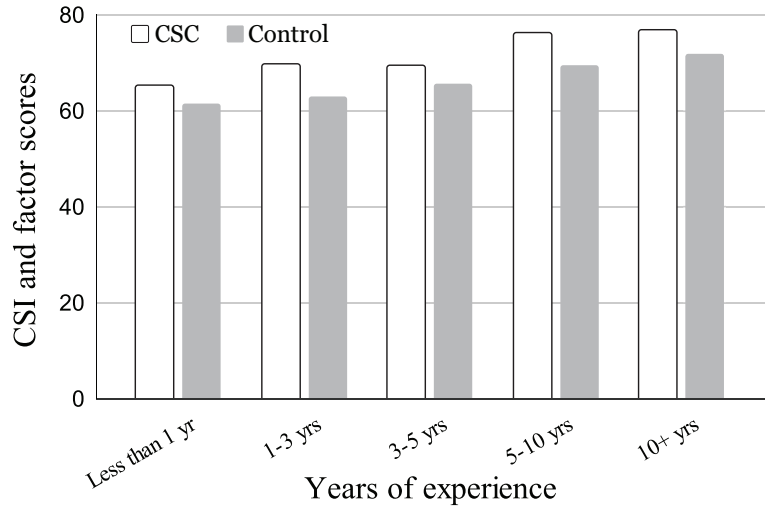}
  \caption{CSI score distribution between groups by years of experience bracket in ``arts \& design'' profession}
 \end{figure}
 


\begin{table}[h]
\begin{center}
\caption{Examples of DCPs generated by participants per tools}
\vspace*{0.5\baselineskip}
\begin{tabular}{p{0.9cm}c p{4cm}p{9cm}} 
\toprule
Tool&\makecell{Des.\\Brief}&$w_1+w_2$&Explanation\\

\midrule
CSC&A&efficient intelligence&My design concept is efficient-intelligence.  It has a sense of intelligence yet it is efficient, not incompetent.  It is efficient, but not ignorance.  In this design, efficient and intelligence can go together.\\
\midrule
CSC&B&harmonious interaction&My design concept is harmonious interaction.  It has a sense of interactivity yet it is harmonious, not incongruous.  It is harmonious, but not incompatibility.  In this design, harmony and interaction can go together.\\
\midrule
Control&B&practical smart&The design is practical and smart, yet is not non-rational. It is practical, not useless.\\
\midrule
Control&A&comfortable utility&This design focuses on comfortable utility, it foregoes the roughness and harshness of typical utility vehicles, yet also avoids the impracticality and compromises of vehicles that put aesthetics over function.\\
\bottomrule
\end{tabular} 
\end{center}
\end{table}


\section{Discussion}
\subsection{Contributions and applications}
The present research showed that our proposed system, CSC, demonstrated its advantage for the goal we aimed to achieve, which is to support the creativity of professional designers. Our research is motivated by an empirical observation on the characteristics of the concept designer task, that is, exploring, identifying, and communicating a design concept, especially with respect to product aesthetics and semantics, as opposed to product features and functionalities. 

We took the example of the automotive industry to describe a case study because the particular characteristics of the design conceptualization task are most apparent in automotive design studios. Yet, these characteristics are common in industrial design in general, where the skills of industrial designers are highly specialized and focused. As \citet{tovey1992intuitive} discussed , the tasks of industrial designers are distinguished from those of their engineering colleagues, and they are specialized in determining the appearance and identity of a product.  However, these practices are traditionally highly empirical, informal, undocumented, and sometimes even secretive due to the strict confidentiality of the industry.  It also tends to be dependent upon the individual experiences of designers.  We modeled this process as generating a compound adjective phrase that best conceptualizes the nature of the design aesthetics and semantics and then constructing a CS to explain them in contrast with an opposing or distinct concept in a quadrant system (Fig.2 (B)). With respect to adjective phrases, as discussed in section 1, these adjective phrases can also be used in design practices on other industries \cite{choo2003effect, khalaj2014comparison, duran2019adjectives}, in computer poetry \cite{toivanen2012corpus}, or even in the advertisements in tourism industry \cite{duran2019adjectives} where the verbal expressions of aesthetics and the semantics of the products matter.  Therefore, the proposed methods can be applied to various design and new product development fields, and are likely to be optimized by introducing domain-specialized corpora for the function \textbf{CalculateWordScore} and the function \textbf{CalculatePhraseScore}.  In regards to the quadrant system, there has been a vague distinction between analysis using a dimensionality reduction on multidimensional scaling (MDS) visualization technique and what designers use as a synthesis tool.  We defined the latter as the CS, translated it into a computational algorithm, and implemented it in an interactive web application so that designers can effectively trace the process in an orderly sequence. Designers know what the important factors of this task are, and evidently, we confirmed that designers value the Exploration factor most in this task. 

In this research, we shed light on concept designers' cognitive activities (Fig. 2-(A)), that were previously indefinite, formalized them (Fig. 2-(B)), translated them into the CSC algorithm (Algorithm 1), and created the interactive user interface (Fig. 3).  The theoretical contribution of the present research is based on a formalization of a designer's concept exploration activities, which involves searching and selecting a compound adjective as a DCP.  This formalization of constructing a CS not only signifies the designers' process of verbal design concept explorations but also leads to subsequent visual creations (Fig.3).

As for a practical implication of the present research, experienced designers seem to have valued the structured process of forming and explaining the DCPs in the experiment. As some of the Weighted Factor Scores of CSI, such as Enjoyment and Immersion, did not seem to matter to them, the participants probably appraised the CSC as to offer them a certain efficiency and effectiveness as a tool for professional use.  This factor evaluation result, with less emphasis on Enjoyment and Immersion, came in with a bit of surprise to us, however, it actually gave us an assurance that we should spend more resources to improve more Exploration factors, which we put many efforts into this research by offering a variety of words and phrases presented on the Explorer, enabled by the ConceptNet knowledge graph.  Also, while a new tool often involves a novelty effect \cite{tsay2020overcoming}, we can claim that the CSC did not introduce a notable novelty effect, as it did not show a significant difference in Enjoyment and Immersion compared with the control condition, where participants used generic tools. 

Regarding the level of expertise of the target audience, about half of creativity support tools (CSTs) have no clear target level of expertise, followed by experts (33\%) and novices (17\%) according to \citet{frich2019mapping}. Our proposed method is clearly targeted for the tasks of experienced designers since we observed the need for support in a particular task. In that sense, as discussed in Fig. 7, though consistent between two groups, there was a tendency indicating that the more experienced the participants were, the more aware they were of the importance of the task, which suggests the necessity or usefulness of this type of creativity support tool. Meanwhile, it can be also said that, if the CSC were successfully emulating the process that an expert designer would use, we should have observed a bottom-up effect for the designers with fewer experiences.  In the future, we may need to promote a method to raise the awareness of process-oriented practices for less experienced designers.

As reported in section 5.1, we observed an order effect between the participants who used CSC first or second. The CSI score for the control tool rated by the participants who used the CSC first substantially dropped. Meanwhile, the CSI score for the CSC tool rated by the participants who used the control tool first practically stayed the same. While this may have been cased by a ``fatigue effect'' on repeated task \cite{bradley1994use} as the tasks given to the participants were fairly complex and time consuming, it is probably a positive sign to see the CSI score which is maintained the same level when used in the second task. 

\subsection{Limitation}
As the focus of this research was to identify and implement the design concept generation process used by designers, that is, creating DCPs and communicating them, into an algorithm for constructing the CS, there are some limitations on the CSC tool.  First, we did not put much emphasis on generating the ``good'' results achieved by some of the related works mentioned in section 2.3., which would have been reflected in the Result Worth Effort factor or Expressiveness of the CSI. For example, the function, \textbf{CalculateWordScore} and the function \textbf{CalculatePhraseScore} could be iterated more in optimizing the algorithms.  In fact, how to optimize suggested words and phrases can be another research question, while the features and characteristics of adjectives can be identified and utilized in order to suggest words and phrases that users can find more creative.  In the course of the research we have examined several potential predictors, such as adjective supersense \cite{tsvetkov2014augmenting}, and psycholinguistic attributes of words \cite{wilson1988mrc}, however, none appeared promising.  There should be more potential predictors we can investigate in the future. Second, the CSC tool focuses on verbal concept design activities and does not involve visual activities at this time.  Although the verbal activity can be a starting point to conceptualize and manipulate the product aesthetics and semantics, designers will engage in visual ideation activities that can be also computationally supported.  We would like to implement this visual process into the system in the future, that is, creating a mood board as we previewed an example (Fig. 4), into the CS.

As for the limitation of our experiment, the participants were not given the identical design brief for each condition, however, we carefully composed both design briefs in consistent manner described in the section 4.1., and deemed those two design briefs equivalent in terms of influences on the results.  Our intention of the experiment was to compare different tools, not to compare the different stimuli though it may leave a room for discussions.  Some may argue it might have been better to set up a between-participants arrangement, yet, we had to take a within-participants arrangement for two reasons: first, we were still trying to test the relative effectiveness between two conditions - the proposed method and what they usually practice in their regular works. Second, recruiting a sufficient body of participants who are in “arts and design” profession for a between-participants arrangement was prohibitively difficult.

We did not ask the participants with control conditions what web search tool(s) they actually used in creating and explaining their DCPs, in order to avoid overloading the participants with additional questions.  Also, those publicly available web search tools do not provide the log data of the participants’ behaviors during the search.  Thus, how many and what queries the participants tried, or how many words the participant actually looked at before deciding $w1$ or $w1-w2$ combinations, for example, is unknown and we were unable to compare such interactions between the conditions. 
 
 Another factor that may affect the CSI results is the graphic user interface.  Our proposed CSC tool did not suggest novelty effect indicated by rather humble Enjoyment and Immersion scores.  However, it can also be said that there is a room to introduce a novelty bias if we adopt a more entertaining graphic user interface.  In the future, we would also like to compare different graphic user interfaces to examine how Enjoyment and Immersion factors would interact with Exploration, Result Worth Effort, or overall CSI scores.

Lastly, despite the fact that the CSI is most commonly used to evaluate creative support tools, we acknowledge that it does not evaluate the “effectiveness” of the tool in terms of how strongly the potential end-users are influenced by the generated design concepts. As there is no one-size-fits-all approach \cite{hewett2005creativity} we may also need to develop our original tool that appropriately measure the effectiveness of the creation while we implement the visual component discussed earlier in this section. 

\section{Conclusion}
We focused on the process of verbal design concept exploration practiced by concept designers, that is, generating and verbally communicating product aesthetics and semantics, and we attempted to implement them into an algorithm. We created a system, Character Space Construction (CSC), that enables concept designers to explore and choose words to create a compound adjective phrase, which we call the Design Concept Phrase (DCP). This is a part of the process of constructing a quadrant system, which we call the Character Space (CS). In a CS, a DCP can be represented in the upper right quadrant, whereas other quadrants can be described as what to stay away from or avoid in a semantic space. Our experiment, which was done with 103 participants using a within-participants arrangement, compared our proposed methods, CSC and the general process designers use, and it was shown that our proposed method, the CSC, has a significantly positive effect using Creativity Support Index (CSI).

\section{list of non-standard abbreviations}
\begin{itemize}
    \item CSC: Character Space Construction
    \item CS: Character Space
    \item DCP: design concept phrase
    \item CSI: Creativity Support Index
    \item MDS: Multidimensional scaling 
\end{itemize}

\section{Acknowledgments}
We appreciate the support of the automotive designers at an anonymous company for cooperating with the preliminary studies on labeling adjectives and adjective phrases.  This work was partially supported by JST, CREST (JPMJCR21D4), Japan.

\section{Data Availability Statement}
The following data, supporting the conclusions of this article, will be made available by the authors, without undue reservation. 
\begin{itemize}
    \item Participants responses to the experiment.
    \item Expert labeling on adjectives and adjective phrases.
    \item The generated design corpus from which we randomly selected the words and phrases for PoS comparisons
    \item Participants responses to the PoS ratings on the preliminary study
    \item The source code of the CSC system
\end{itemize}
The following datasets presented in this study can be found in online repositories. The names of the repositories can be found in the article.
\begin{itemize}
    \item The ConceptNet API
    \item The word embedding vectors of the ConceptNet Numberbatch
    \item EnTenTen15 English web corpus
\end{itemize}

\section{Conflict of Interest Statement}
Author Shin Sano was employed by Toyota Motor Corporation and its North American subsidiary, Calty Design Research, Inc. between April 1992 and July 2007. Author Shin Sano was hired as a design consultant between January 2015 and July 2020 for the Concept-i project, which is introduced as an industry case study in the introduction section of this paper. This research is independent of any part of that project and the authors received no funding for the cost of this research. The remaining author declared that the research was conducted in the absence of any commercial or financial relationships that could be construed as a potential conflict of interest.

\bibliographystyle{unsrtnat}
\bibliography{Sano_FI}

\end{document}